\documentclass[aps,twocolumn,superscriptaddress,floatfix,preprintnumbers,nofootinbib]{revtex4-1}
\usepackage[utf8]{inputenc}
\usepackage{amsmath}
\usepackage{amssymb}

\usepackage{enumerate}
\usepackage{amsmath}
\usepackage{tikz}
\usepackage[compat=1.1.0]{tikz-feynman}
\usepackage{feynmf}
\usepackage{slashed}
\usepackage{braket}
\usepackage[toc,page]{appendix}
\usepackage{url}
\usepackage{natbib}
\usepackage{graphicx}
\usepackage[pdftex,bookmarks,linktocpage,pdfpagelabels,plainpages=false,hyperfigures,linkcolor=blue,citecolor=blue]{hyperref}

\hypersetup{colorlinks=true}

\usepackage{mathrsfs,amssymb}  
\usepackage{cancel}
\usepackage[normalem]{ulem}
\usepackage[perpage]{footmisc}
\usepackage[capitalize]{cleveref}
\begin{document}

\begin{flushright}
\preprint{MI-HET-778,~~FERMILAB-PUB-22-453-T} 
\preprint{NUHEP-TH/22-04} 

\end{flushright}

\author{Vedran~Brdar}
\affiliation{
Fermi National Accelerator Laboratory, Batavia, IL, 60510, USA}
\affiliation{Northwestern University, Department of Physics \& Astronomy, 2145 Sheridan Road, Evanston, IL 60208, USA}

\author{Bhaskar~Dutta}
\affiliation{Mitchell Institute for Fundamental Physics and Astronomy,
Department of Physics and Astronomy, Texas A\&M University, College Station, TX 77843, USA}

\author{Wooyoung~Jang}
\affiliation{Department of Physics, University of Texas, Arlington, TX 76019, USA}

\author{Doojin~Kim}
\affiliation{Mitchell Institute for Fundamental Physics and Astronomy,
Department of Physics and Astronomy, Texas A\&M University, College Station, TX 77843, USA}

\author{Ian~M.~Shoemaker}
\affiliation{Center for Neutrino Physics, Department of Physics, Virginia Tech, Blacksburg, VA 24061, USA}

\author{Zahra~Tabrizi}
\affiliation{Northwestern University, Department of Physics \& Astronomy, 2145 Sheridan Road, Evanston, IL 60208, USA}

\author{Adrian~Thompson}
\affiliation{Mitchell Institute for Fundamental Physics and Astronomy,
   Department of Physics and Astronomy, Texas A\&M University, College Station, TX 77843, USA}

\author{Jaehoon~Yu}
\affiliation{Department of Physics, University of Texas, Arlington, TX 76019, USA}
\affiliation{Neutrino Department, CERN, Geneva, Switzerland}

\title{BSM Targets at a ``Target-less DUNE''}

\begin{abstract}
In this Letter we demonstrate that a future accelerator-based neutrino experiment such as DUNE can greatly increase its sensitivity to a variety of new physics scenarios by operating in a mode where the proton beam impinges on a beam dump. 
We consider two new physics scenarios, namely light dark matter (LDM) and axion-like particles (ALPs) and show that by utilizing a dump mode at a DUNE-like experiment, unexplored new regions of parameter space can be probed with an exposure of only 3 months with half of its expected initial beam power. Specifically, target-less configuration of future high intensity neutrino experiments will probe the parameter space for thermal relic DM as well as the QCD axion (DFSZ and KSVZ). The strength of such configuration in the context of new physics searches stems from the fact that the neutrino flux is significantly reduced compared to that of the target, resulting in much smaller backgrounds from neutrino interactions.
We have verified this in detail by explicitly computing neutrino fluxes which we make publicly available in order to facilitate further studies with a target-less configuration.
\end{abstract}

\maketitle

\noindent {\bf Introduction.} 
The remaining unknowns in the standard three flavor oscillation paradigm are expected to be measured by the next generation neutrino experiments among which  DUNE 
\cite{DUNE:2020lwj,DUNE:2020ypp,DUNE:2020mra,DUNE:2020txw}, Hyper-Kamiokande \cite{Hyper-KamiokandeWorkingGroup:2014czz} and JUNO \cite{JUNO:2015zny} stand out. While determination of the mass ordering, the octant of the atmospheric mixing angle and the level of CP violation in the lepton sector are indeed the main priority, it should be stressed that such experiments can also be utilized in the context of various new physics searches. Specifically, in addition to new physics realizations (\emph{e.g.} light eV-scale sterile neutrinos \cite{Dasgupta:2021ies}, non-standard interactions \cite{Farzan:2017xzy} and ultra-light dark matter (DM) \cite{Krnjaic:2017zlz,Brdar:2017kbt}) that could make a direct impact to the measured oscillations, it was shown particularly that such experiments will also be very competitive in complementary beyond the Standard Model (BSM) tests (see for instance section 7 in 
\cite{Arguelles:2019xgp,DUNE:2020fgq} and references therein). Focusing on a liquid Argon near detector, let us stress 
searches for LDM \cite{DeRomeri:2019kic,Celentano:2020vtu,Breitbach:2021gvv}, axion-like particles (ALPs) \cite{Brdar:2020dpr,Kelly:2020dda}, new vector bosons~\cite{Ballett:2019xoj,Berryman:2019dme,Dev:2021qjj,Capozzi:2021nmp,Chauhan:2022iuh}, heavy neutral leptons~\cite{Ballett:2018fah,Ballett:2019bgd,Berryman:2019dme,Coloma:2020lgy,Atkinson:2021rnp,Abdullahi:2022jlv} and model-independent BSM searches in the context of the Standard Model Effective Field Theory~\cite{Falkowski:2018dmy}. 

One of the main challenges in performing such BSM searches at neutrino experiments is that the corresponding signals often appear in the detector with a very similar signature as  neutrino interactions, and hence the neutrinos act as the main source of  background. One of the proposals to reduce the effect of such background is to place the detector off-axis, the case in which the ratio between the BSM signal and the neutrino background increases~\cite{Coloma:2015pih,DeRomeri:2019kic}. The challenge is that by going off-axis, the signal statistics also decreases significantly, and therefore, several years of data taking is necessary to obtain competitive constraints in the BSM parameter space.  

In this letter, we propose a complementary configuration for a future DUNE-like experiment, inspired by the method performed at the MiniBooNE beam-dump experiment \cite{MiniBooNEDM:2018cxm}. The idea is that if the proton beam can be directed away from the target (and hence directly impinge on the beam dump), the neutrino flux will be greatly reduced, as charged mesons get absorbed in the dump before they decay. Further, this configuration provides additional materials for photon flux to increase; hence the effective distance between the production source at the dump and the detector decreases, which increases the solid angle coverage and the number of  BSM particles entering the detector (for the case of DUNE the BSM flux will be roughly increased by a factor of $4$). While the long term usage of the target-less configuration would clearly not allow for precise measurement of neutrino parameters, in this work we will demonstrate that for several new physics scenarios large and hitherto unconstrained regions of parameter space can be probed with only three months of data taking at half the expected initial beam intensity of DUNE. The relatively short duration needed here will therefore not dramatically influence neutrino program at DUNE-like experiments where the remaining unknowns in the standard three-neutrino paradigm are expected to be obtained.

\medskip

\noindent {\bf Target-less Configuration.}
In a typical accelerator neutrino experiment, neutrinos are produced from the decays of the secondary charged mesons generated by the proton interactions in the target. The surviving hadrons are captured in the hadron absorber or a beam dump. In this work, however, we consider a target-less configuration for which the proton beam directly makes collisions with the beam dump. To estimate the sensitivity of this configuration for different physics cases, we first model the fluxes of neutrinos, which is the main source of background to BSM physics signals, and the progenitor particles ($\pi^0$, $\eta^0$, $e^\pm$, $p$), which is the source of the LDM and/or ALP signals. 
We have simulated the fluxes using GEANT4~\cite{Agostinelli:2002hh} with the  \texttt{QGSP\_BIC\_AllHP} physics list for the hadronic reactions and \texttt{G4EmStandardPhysics} for the electromagnetic interactions. For the case of DUNE we have used a cube-shaped aluminum core of $4\times 4\times4~$m$^{3}$ for the dump and we have assumed a 120-GeV proton beam impinging on it.

\begin{figure}
  \centering
  \includegraphics[width=0.45\textwidth]{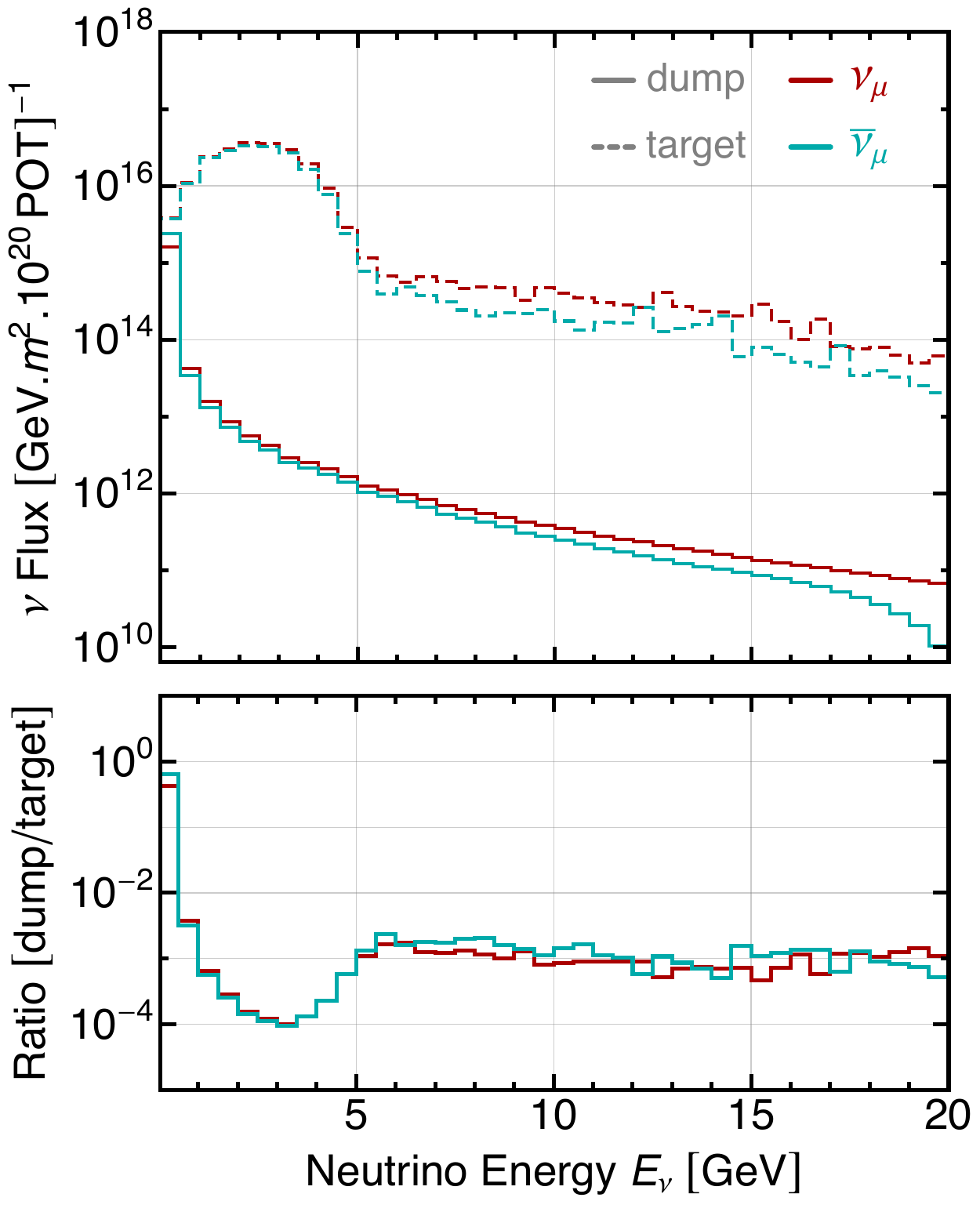}
\caption{ The target-less configuration fluxes (solid) obtained in this work and neutrino fluxes for the target (dashed), adopted from Ref.~\cite{DUNE:2020ypp}. Here, the fluxes of $\nu_\mu$ and $\bar\nu_\mu$ in the target mode are the results obtained in the forward horn current mode and the reverse horn current mode, respectively. The red and cyan curves show the fluxes of $\nu_\mu$ and $\bar\nu_\mu$, respectively. Absolute fluxes for target and dump case are shown in the upper panel while their ratio is shown in the lower panel.
  }
  \label{fig:fluxDumeVsTarget}
\end{figure}

The resulting flux of neutrinos in the target-less mode as a function of the neutrino energy is shown in the top panel of  \cref{fig:fluxDumeVsTarget} (solid curves), where the $\nu_\mu$ and $\bar\nu_\mu$ fluxes are shown in red and blue, respectively. To demonstrate the potential of the target-less mode in suppressing the neutrino fluxes, we have also shown the expected neutrino fluxes for the target case in dashed curves, taken from Ref.~\cite{DUNE:2020ypp}. The bottom panel shows the ratios of the fluxes in the target-less compared to the target mode. 
 The peaks for both $\nu_\mu$ and $\bar\nu_\mu$ in the first bin for the target-less configuration is due to the charged meson decay-at-rest, and the dip around 3 GeV in the ratio can be understood as the effect of the focusing horns which is not present in the target-less configuration. One can see that an overall flux suppression of $10^{-3}-10^{-4}$ is achievable for the dump case, which demonstrates the potential of the DUNE target-less configurations to significantly suppress the neutrino background. 
In order to verify these results, we have performed a separate simulation based on the MiniBooNE beam dump assumptions and have found general agreement with the neutrino fluxes in Ref.~\cite{MiniBooNE:2017nqe}.  

As a final remark here let us emphasize that the DUNE dump is designed to survive two consecutive accidental deposition of 2.4 MW proton beams, effectively. To estimate the sensitivity of the target-less mode to different BSM cases, we consider two scenarios: $(i)$ a realistic case assuming the beam power is 0.6 MW with a run time of 3 months, and $(ii)$ an optimistic case with a beam power of 1.2 MW and a run time of 1 year. We show the results of these two cases throughout the paper. 

\medskip

\noindent {\bf Benchmark Physics Cases.}
In what follows we demonstrate the potential of a target-less configuration at DUNE-like experiments in BSM searches by focusing on two new physics scenarios:  Light Dark Matter (LDM) and Axion-Like Particles (ALPs).\\

{\emph{Light Dark Matter.}} In recent years, significant attention has been paid to studying light MeV-GeV scale dark matter at neutrino experiments~\cite{DeRomeri:2019kic,Celentano:2020vtu,Breitbach:2021gvv,Naaz:2019pvs}. These studies are done by utilizing the intense proton beam at these facilities which strike the target, producing charged and neutral mesons; latter may then produce DM particles and this typically goes via meson decay into a final state containing dark photon which couples to DM particles. The DM can then travel to the detector and produce a signal by elastic scattering on electrons. In this work we consider the following dark-sector model by augmenting the SM with a complex scalar DM $\phi$, charged under $U(1)_D$ symmetry, which communicates with the SM through a dark photon  $A'$ (the $U(1)_D$ gauge boson), described by the Lagrangian \cite{Batell:2009di,deNiverville:2011it,deNiverville:2012ij,Izaguirre:2013uxa}
\begin{equation}
\mathscr{L}_{\rm{DM}}\supset -\frac{1}{2}\epsilon F_{\mu\nu}F'^{\mu\nu}+ \frac{1}{2} m_{A'}^2 A'_\mu A'^\mu + |D_\mu \phi|^2-m_\phi^2|\phi|^2 \,,
\end{equation}
where $\epsilon$ is the kinetic mixing, $F_{\mu\nu}$ and $F'_{\mu\nu}$ are the electromagnetic and dark photon field strength tensors, respectively, $m_{A'}$ and $m_\phi$ are the dark photon and DM masses, and $D_\mu$ is the covariant derivative corresponding to $U(1)_D$ with the coupling $g_D$ and dark-sector fine structure constant $\alpha_D\equiv g_D^2/4\pi$.
Given this Lagrangian, dark photons can be produced through various mechanisms. Among them, dominant production channels include the following ones: $(i)$ neutral mesons (e.g., $\pi^0$, $\eta$) can decay into a SM photon and a dark photon $A'$ through the kinetic mixing, i.e., $\pi^0/\eta \to \gamma \, A^\prime$, $(ii)$ the incoming proton beam can scatter off a nucleus, emitting a dark photon, i.e., $p \, N \to p \, N \, A^\prime$, and $(iii)$ positrons arising in the secondary electromagnetic showering process of the charged particles created by the beam collision can produce dark photons resonantly, together with electrons in target atoms, i.e., $e^+ \, e^- \to A^\prime$. We use GEANT4 with the \texttt{QGSP\_BIC\_ALLHP} physics list to simulate sample fluxes of the progenitor particles ($\pi^0$, $\eta^0$, $e^\pm$, $p$), which can then source the DM flux, $\mathcal{F}_{DM}$.

\begin{figure}[t!]
  \centering
  \includegraphics[width=0.47\textwidth]{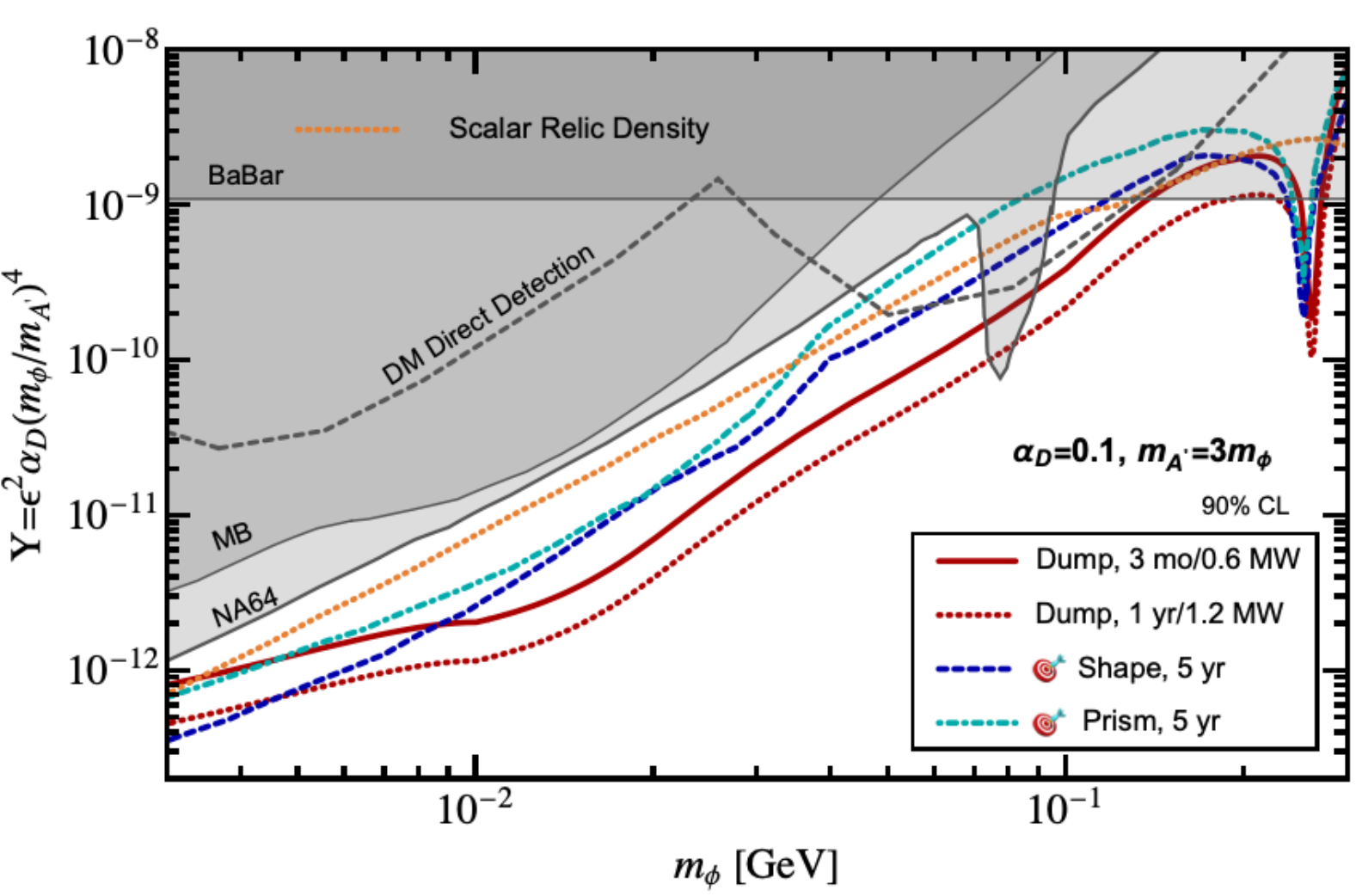}
\caption{LDM sensitivity for target-less run of 3 months-0.6 MW beam (1year-1.2 MW beam) shown in solid (dashed) red curve,  compared to the DUNE target configurations discussed in the literature, namely the shape analysis (blue dashed curve) of Ref.~\cite{Breitbach:2021gvv} and the PRISM analysis of Refs.~\cite{Breitbach:2021gvv,DeRomeri:2019kic} (cyan dot-dashed curve). The ``target'' symbol in the legend denotes the target mode measurements. The thermal relic density for scalar DM reproducing the observed DM abundance is shown in orange. The existing constraints from various experiments are shown in gray shaded regions. See text for more details.}
  \label{fig:DM}
\end{figure}

Once produced, $A^\prime$ immediately decays to a DM pair, and the DM can manifest itself as scattering on e.g., electrons, $\phi+e^-\rightarrow \phi+e^-$. The differential cross section with respect to the electron recoil energy, $E_r$, is given by \cite{deNiverville:2011it}
\begin{eqnarray}
    \frac{d\sigma_{\phi e}}{dE_r}=4\pi\epsilon^2\frac{\alpha_D \alpha_{\rm{EM}}}{p_\phi^2}\frac{2E_\phi m_e(E_\phi-E_r)-m_\phi^2 E_r}{(2m_eE_r+m_A'^2)^2}   \,,
\end{eqnarray}
where $m_e$ is the electron mass, $\alpha_{\rm EM} \simeq 1/137$ is the QED fine structure constant and $E_\phi$ ($p_\phi$) is the DM energy (momentum).  
The expected number of DM signal  $N_{\rm DM}$ at the detector is given by
\begin{equation}
    N_{\rm DM}=N_e \int dE_\phi \frac{d\mathcal{F}_{\rm DM}}{dE_\phi} \sigma_{\phi e}(E_\phi)\,, 
\end{equation}
where $N_e$ is the number of electron targets inside the detector fiducial volume, for which we have assumed 50-t of liquid argon (LAr). We have set the threshold energy to 30~MeV~\cite{DUNE:2020ypp}. 

In~\cref{fig:DM} we have shown the target-less configuration's sensitivity to LDM using the dimensionless parameter $Y \equiv \epsilon^2 \alpha_D (m_{\phi}/m_{A'})^4$, for dark sector coupling we chose $\alpha_{D} = 0.1$, and assumed that the DM mass, $m_\phi$, is three times smaller than the dark photon mass $m_{A'}$. 
For the realistic case of 0.6~MW beam power and a run time of 3~months (optimistic case of 1.2-MW in 1~year), we require $N_{\rm DM}=3.1 ~(8.0)$ to obtain 90\% CL sensitivity projections. Note that neutrino fluxes in the target-less configuration, shown in~\cref{fig:fluxDumeVsTarget}, yield 0.5 (4) neutrino-electron scattering background events and that information has been included in setting sensitivity projections. The red curves in~\cref{fig:DM} represent the sensitivity projections for the two aforementioned study cases. One can see that for $m_\phi\gtrsim 10~$MeV a run of only 3 months yields stronger sensitivity projections than other DUNE configurations with a run time of 5 years, e.g. a careful bin by bin analysis using the on-axis detector~\cite{Breitbach:2021gvv} (dashed blue curve), or using the PRISM concept to suppress the background~\cite{Breitbach:2021gvv,DeRomeri:2019kic} (dot-dashed cyan curve).

One can infer that the reach in the target-less realization also ``exceeds'' the orange line corresponding to the DM abundance in accord with present observations. Hence, a run of only 3 months will have the ability to probe thermal relic LDM up to $m_{\phi} \simeq 200$ MeV ($m_{A'} \simeq 600$ MeV). Given the reach of the target-less projection relative to the DM abundance line, scenarios in which thermal relic is a subdominant DM component are also testable. Further, note that asymmetric DM models generically require larger than thermal relic cross sections~\cite{Graesser:2011wi}, and hence these models are more testable at a DUNE-like experiment compared to (symmetric) thermal relic DM. We have also shown the existing constraints in shaded grey regions from the MiniBooNE~\cite{MiniBooNEDM:2018cxm}, BaBar~\cite{BaBar:2017tiz} and NA64~\cite{Andreev:2021fzd} experiments. Current direct detection experiments, for example XENON1T, XENON10 and SENSEI (shown in gray dashed curve) also cover the same parameter space, and are marginally better than NA64 only for masses larger than $m_\phi\gtrsim 100~$MeV, while for masses around a few MeV the sensitivity is only an order of magnitude better than BaBar~\cite{SENSEI:2020dpa}. \\

{\emph{Axion-Like Particles.}}
Axions and axion-like particles (ALPs) span a wide range of well-motivated models, namely those that solve the strong CP problem (QCD axions) and those that feature the ALP as a viable DM candidate (see Refs.~\cite{Chadha-Day:2021szb, Marsh:2015xka} for a review), but otherwise arise more generally from the zero modes of string theory models~\cite{Arvanitaki:2009fg}. A vast effort is being made into searching for ALPs using their couplings with the SM particles at the current and future experiments:~\cite{Zioutas:1998cc,Anastassopoulos:2017ftl,Irastorza:2013dav,Kahn:2016aff,Salemi:2019xgl,Asztalos:2001tf,Du:2018uak,JacksonKimball:2017elr,Brubaker:2016ktl,Droster:2019fur,Abdullahi:2022jlv,Spector:2019ooq,Abdullahi:2022jlv,Melissinos:2008vn,DeRocco:2018jwe,Liu:2018icu,Obata:2018vvr,Feng:2018noy,Berlin:2018bsc,Akesson:2018vlm,Volpe:2019nzt,Dusaev:2020gxi,Banerjee:2020fue,Berlin:2018pwi,Alekhin:2015byh,Bonivento:2019sri,Dent:2019ueq,AristizabalSierra:2020rom,Aprile:2020tmw,Dent:2020jhf,PhysRevD.101.052008,Fu_2017,Dent:2020qev,Balkin:2021jdr}.

To investigate the ALP parameter space, we will focus on a generic model where ALP couples to an electron-positron pair and photons as described by interaction terms in the Lagrangian of the form 
\begin{equation}
\mathcal{L}_{\rm int} ~\supset~ -i\,g_{ae}\,a\,\bar{\psi}_e \, \gamma_5\, \psi_e\,-\frac{g_{a\gamma}}{4}\,a\,F_{\mu\nu}\tilde{F}^{\mu\nu}\,,
\label{eq:alp}\end{equation}
where $a$ denotes the ALP field and $\psi_e$ is the electron field. $F^{\mu\nu}$ ($\tilde{F}^{\mu\nu}$) is the electromagnetic (dual) field strength. The couplings $g_{ae}$ and $g_{a\gamma}$ are assumed to be independent for this analysis. These operators capture, for example~\cite{DiLuzio:2020wdo, GrillidiCortona:2015jxo}, phenomenology of  Dine-Fischler-Srednicki-Zhitnitsky (DFSZ) type models~\cite{Zhitnitsky:1980tq,DINE1981199,Dine:1981rt,Dine:1982ah} of QCD axions which feature couplings to SM fermions generated by the dynamics of an extended Higgs sector and PQ field after spontaneous symmetry breaking. 
 Alternatively, Kim-Shifman-Vainshtein-Zakharov (KSVZ)~\cite{PhysRevLett.43.103, SHIFMAN1980493} variants can lead to operators like $a F \tilde{F}$ through loop diagrams of heavy color-charged fermions. We do not, however, restrict ourselves to the parameter space of traditional QCD axion models, permitting any phenomenologically accessible parts of the parameter space of \cref{eq:alp}.

Given the interaction Lagrangian in \cref{eq:alp}, we consider multiple ALP production channels that arise from this operator. To compute production of ALPs via $g_{ae}$ we employ the photons, electrons, and positrons produced from secondary interactions of protons incident on the dump volume, again using GEANT4 with the \texttt{QGSP\_BIC\_ALLHP} physics list. The explicit production channels are Compton scattering ($\gamma \, e^- \to e^- \, a$) from $\gamma$ incident on electrons at rest, associated production ($e^+ \, e^- \to \gamma \, a$) and resonant production ($e^+ \, e^- \to a$) both arising from positron annihilation on electrons at rest, and finally, ALP-Bremsstrahlung ($e^\pm \, N \to e^\pm  \, N \, a$) producing ALPs from braking radiation of $e^\pm$ through the atomic targets $N$ in the dump material; see refs.~\cite{Capozzi:2021nmp, CCM:2021lhc, Nardi:2018cxi, PhysRevD.34.1326} for other uses of these channels and the analytical techniques therein. 
ALPs coupling to photons ($g_{a\gamma}$ coupling) are produced via the coherent Primakoff process ($\gamma \, N \to a \, N$) due to $\gamma$ incident on atomic targets in the dump material.

From the detection side, we consider several signatures of ALPs in the DUNE-like detector. ALPs coupling to electrons through $g_{ae}$ could decay to an electron-positron pair with the well-known decay width, which in conjunction with the ALP energy, fixes the decay length. Also, an $e^+ \, e^-$ final state can be produced from ALPs undergoing external pair production (similar to SM photon pair production) $a \, N \to e^+ \, e^- N$ via interaction with the strong electric field of the nucleus. Lastly, ALPs could also yield the signal by scattering in the detector via  $a \, e^- \rightarrow \gamma \, e^-$, which produces a soft, slightly off-forward photon as well as a hard and forward-going electron. For ALPs coupling to photons through $g_{a\gamma}$, we consider the ALP decays to two photons $a \to \gamma \gamma$ in the LAr volume as well as inverse-Primakoff scattering, $a \, N \to \gamma \, N$. 
\begin{figure}[h]
    \centering
    \includegraphics[width=0.47\textwidth]{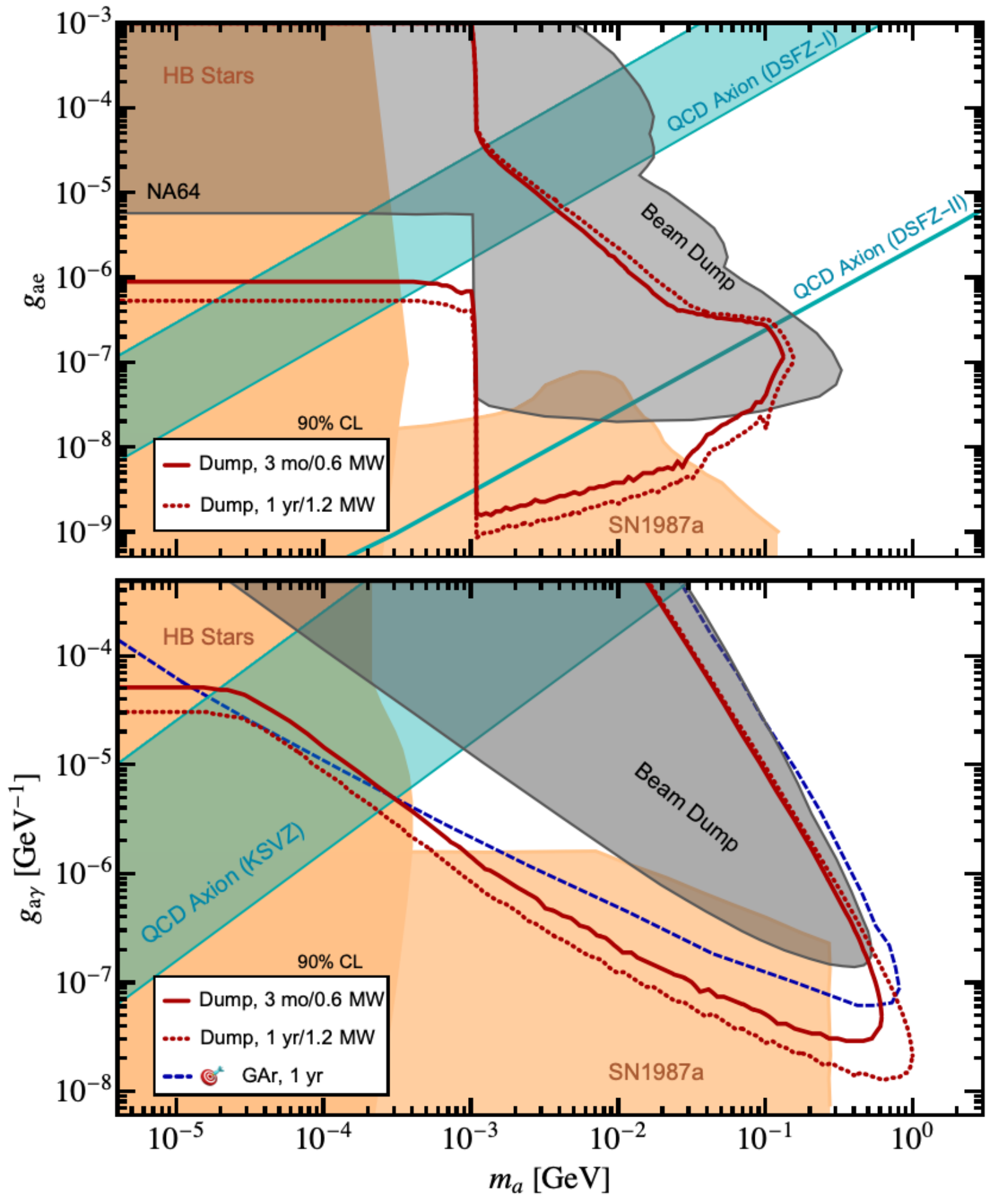}
    \caption{The target-less configuration sensitivity to the ALP-electron and ALP-photon coupling using 3 months-0.6 MW power (1 year-1.2 MW power) shown in red solid (dashed) curves. {\emph{Top panel}:} For $g_{ae}$, Compton scattering, associated production, resonant production, and bremsstrahlung production of ALPs are considered via $\gamma$ and $e^\pm$ scattering in the dump environment. Detection channels include ALP decay to $e^+e^-$ for $m_a > 2m_e$ as well as external pair conversion and inverse Compton scattering in the detector (see text for details). Also displayed are the DFSZ model preferred regions, where DFSZ-I and DFSZ-II are sub-models differentiated by the relationship between the axion-lepton and axion-quark couplings. {\emph{Bottom panel}:}
    For $g_{ae}$, the production occurs via Primakoff scattering process. The detection channels involve the inverse-Primakoff scattering process and two photon final states from the decay. KSVZ model preferred regions are shown as well. The blue dashed curve is the 1 year target result using the gas detector in Ref.~\cite{Brdar:2020dpr}. The ``target'' symbol in the legend denotes this target mode measurement. 
    }
    \label{fig:alp_sensitivity}
\end{figure}
At the target case the main source of background for ALP decay is the production of NC-$\pi^0$ which can decay into a pair of photons. Using a gaseous argon detector (GAr) at a DUNE-like experiment one can veto most of this background by the hadronic activity as well as employing kinematic variables~\cite{Brdar:2020dpr}. The situation at the LAr detector is vastly different, and such backgrounds have never been studied before. Here we stress again that for target-less configuration at a DUNE-like experiments all such backgrounds are practically negligible, since the original neutrino flux is suppressed, as shown in \cref{fig:fluxDumeVsTarget}.


In Fig.~\ref{fig:alp_sensitivity}, we show  sensitivity of a target-less DUNE-like experiment run to the $g_{ae} - m_a$
parameter space (top) and $g_{a\gamma}-m_a$ parameter space (bottom) for our two benchmark scenarios (3 months-0.6 MW power
case is shown in solid and 1 year-1.2 MW realization is shown in dashed red). Existing laboratory limits on
$g_{ae} - m_a$~\cite{Bjorken:1988as,Andreas:2010ms,Bechis:1979kp,NA64:2021aiq,Gninenko:2017yus,Andreev:2021fzd,Riordan:1987aw,Bross:1989mp} 
and $g_{a\gamma}-m_a$~\cite{Blumlein:1991xh,Blumlein:1990ay,Jaeckel:2015jla} are shown in gray while the astrophysical probes from supernovae~\cite{PhysRevD.104.103007, Payez:2014xsa}
and stellar cooling~\cite{Dolan:2021rya,Hardy:2016kme, Carenza:2020zil} are also shown (shades of orange). Regarding supernovae, we also point out \cite{Lucente:2020whw} 
featuring more developments in the calculations of the supernova limit as well as Refs.~\cite{Caputo:2022mah,Caputo:2021rux} where the authors discuss ruling out the
so-called ``cosmological triangle'' \cite{Brdar:2020dpr}. 

As one can see from the figure, a 3 month exposure in the target-less realization can result in significantly improved sensitivity to ALPs,
and test viable QCD axion models of both the DFSZ and KSVZ benchmark scenarios. For the  $g_{a\gamma}$ coupling this can even improve over the 1 year study at the GAr detector for most of the parameter space.
We also notice the overlap of the sensitivity with supernova limits, and given the aforementioned supernovae uncertainties, the target-less realization of DUNE will clearly be able to make statements about the validity of such astrophysical considerations.

\medskip

\noindent {\bf Conclusions.}
In this work we have considered a scenario in which proton beam from a DUNE-like experiment impinges directly on the beam dump. We have shown, via explicit calculation, that neutrino fluxes in such configuration are rather small, guaranteeing effectively background-free BSM searches. We have investigated a simplified light dark matter (LDM) model as well as photophilic and photophobic ALPs in this context and have shown that by running in the dump mode for only 3 months with a beam power of 0.6 MW, presently unconstrained portions in the parameter space can be probed.  It should be stressed that this target-less configuration can probe high-priority new physics targets including DM interactions consistent with the thermal relic hypothesis for the observed DM density, as well as DFSZ and KSVZ models of the QCD axion.

\medskip

\noindent{\bf{Acknowledgements.}}
BD and AT acknowledge support from the U.S. Department of Energy (DOE) Grant DE-SC0010813.
The work of DK is supported by DOE under Grant No. DE-FG02-13ER41976/DE-SC0009913/DE-SC0010813.  
The work of IMS is supported by DOE under the award number DE-SC0020250. 
The work of ZT is supported by the Neutrino Theory Network Program Grant No. DE-AC02-07CHI11359 and the U.S. Department of Energy under the award number DE-SC0020250. 
The work of AB, WJ and JY is supported by the U.S. Department of Energy under Grant No. DE-SC0011686.
JY thanks the support of the CERN neutrino department during his stay in which majority of this work is performed.
Fermilab is operated by the Fermi Research Alliance, LLC under contract No. DE-AC02-07CH11359 with the United States Department of Energy. Portions of this research were conducted with the advanced computing resources provided by Texas A\&M High Performance Research Computing. This research was supported in part by the National Science Foundation under Grant No. PHY-1748958.

Disclaimer: We speak for ourselves, not on behalf of the DUNE Collaboration. This work is based on our ideas, our calculations, and publicly available information.

\bibliography{main}

\end{document}